\documentclass[sigplan,11pt,nonacm=true]{acmart}
\usepackage{graphicx} 
\usepackage{caption}
\usepackage{amsmath}
\usepackage{listings}
\usepackage{multirow}
\usepackage{color}

\title{Assessing the Code Clone Detection Capability of Large Language Models}

\author{Zixian Zhang}
\affiliation{%
  \institution{School of Computer Science, University of Galway}
  \country{Ireland}
}
\email{z.zhang15@universityofgalway.ie}
\authornote{Also affiliated with CRT-AI, the Science Foundation Ireland Centre for Research Training in Artificial Intelligence.}

\author{Takfarinas Saber}
\affiliation{%
  \institution{School of Computer Science, University of Galway}
  \country{Ireland}
}
\email{takfarinas.saber@universityofgalway.ie}
\authornote{Also affiliated with Lero, the Science Foundation Ireland Research Centre for Software.}

\begin{abstract}
This study aims to assess the performance of two advanced Large Language Models (LLMs), GPT-3.5 and GPT-4, in the task of code clone detection. The evaluation involves testing the models on a variety of code pairs of different clone types and levels of similarity, sourced from two datasets: BigCloneBench (human-made) and GPTCloneBench (LLM-generated). Findings from the study indicate that GPT-4 consistently surpasses GPT-3.5 across all clone types. A correlation was observed between the GPTs' accuracy at identifying code clones and code similarity, with both GPT models exhibiting low effectiveness in detecting the most complex Type-4 code clones. Additionally, GPT models demonstrate a higher performance identifying code clones in LLM-generated code compared to humans-generated code. However, they do not reach impressive accuracy. These results emphasize the imperative for ongoing enhancements in LLM capabilities, particularly in the recognition of code clones and in mitigating their predisposition towards self-generated code clones--which is likely to become an issue as software engineers are more numerous to leverage LLM-enabled code generation and code refactoring tools.

\end{abstract}


\begin{document}
\maketitle

\section{Introduction}
\label{sec:introduction}

The detection of code clones as well as the assessment of code similarity hold considerable importance in the field of software engineering. Identifying and managing code clones is crucial for maintaining code quality and integrity~\cite{de2018systematic}. It aids in reducing redundancy, preventing bugs, and ensuring consistency across the codebase. Furthermore, the assessment of code similarity helps detect code leaks/plagiarism~\cite{10.1109/imitec52926.2021.9714688,10.36548/jaicn.2020.3.005} and improve automatic code generation~\cite{tao2022assessing,tao2022multi,tao2023many}. Extensive research efforts have been dedicated to code clone detection methodologies over the decades. Despite these efforts, identifying semantic clones—code pairs characterized by low textual similarity—remains a big challenge. 

In the wake of artificial intelligence (AI) revolution, scholars have sought solutions through AI-based methods. Increasingly, Machine Learning (ML) and Deep Learning (DL) techniques have been harnessed to unearth code snippets bearing semantic or syntactic similarities, utilizing tools such as Convolutional Neural networks (CNNs)~\cite{sheneamer2021effective, li2020deep}, Recurrent Neural Networks (RNNs)~\cite{yuan2020local}, Graph Neural Networks (GNNs)~\cite{dai2023study, liu2022low}, and transformers \cite{zhang2023efficient}. Nonetheless, there have been no studies that specifically explore the application of Large Language Models (LLMs) in detecting code clones.

Recent years have witnessed a rapid advancement in the field of Large Language Models (LLMs), characterized by their increasing complexity and capabilities. This evolution led to a wide range of new applications in various domains including healthcare~\cite{lyu2023translating, gilson2023does}, machine translation~\cite{gu2023linguistically, zhu2023multilingual}, and recommendation systems~\cite{wang2023zero, dai2023uncovering}. In comparison to other domains addressed by LLMs, software engineering presents a unique set of challenges. The precision required in coding, coupled with the necessity for logical consistency, error-free execution, performance, quality, maintainability, and evolution, makes software engineering a significantly more complex domain for LLMs.

Several studies have investigated the performance of LLMs in some aspects of software engineering. These studies have endeavored to explore and evaluate the efficacy of LLMs in understanding and generating code \cite{tao2023program, destefanis2023preliminary, 10.48550/arxiv.2305.11837}, generation of code documentation \cite{khan2022automatic} or repairing code \cite{10.48550/arxiv.2304.02195}. To the best of our knowledge, the study by Wang et al. \cite{wang2023empirical} is the only to evaluate the performance of LLMs on code clone detection. However, this study is limited to ChatGPT, and does not explore diverse types of code clones or multiple sources of datasets (including real-world code clones and clones generated by the GPT model itself).

To fill this gap, in this paper, we aim to systematically evaluate the capabilities of two LLM models (i.e., GPT-3.5 and GPT-4) in identifying code clones. Furthermore, as Integrated Development Environments (IDEs) become integrated with LLM tools (e.g., GPT-4 and Microsoft Copilot) and software engineers are more numerous to leverage their capabilities for various software engineering tasks (e.g., code generation and refactoring), we would like to assess whether there is a difference between the performance of LLM models at identifying human-generated code clones in comparison to LLM-generated code clones.

Our paper aims at answering the following Research Questions (RQs):
\begin{itemize}
    \itemindent=-13pt
    \item \textbf{RQ1: } How does the performance of GPT-3.5 compare to the performance of GPT-4 across various code clone types and code similarity levels?
    \item \textbf{RQ2: } Do GPT models exhibit different performances when assessing human-generated versus LLM-generated code clones? If so, how do these differences manifest between GPT-3.5 and GPT-4?
\end{itemize}

The rest of the paper is organized as follows. Section~\ref{sec:background} summarises the background of our study. Section~\ref{sec:approach} describes the approaches we used to select data and design GPT prompt. In Section~\ref{sec:results}, we analyze the performance of GPT through the data we selected. Finally, Section~\ref{sec:conclusion} concludes the paper.

\section{Background}
\label{sec:background}

In this section, we describe the background of our study in two parts: Code Clones and Generative Pretrained Transformers.

\subsection{Code Clones}
A sequence of source code is known as a code fragment that is identical or very similar to another code fragment, this pair is known as clone pairs. These pairs are similar in terms of functionality, structure, or both. These pairs can be found within the same file, across different files in a single project, or even across projects. There are various clone types based on clone pair similarity. We use the definition of clones which has been widely accepted by many scholars \cite{roy2009comparison}:

\begin{itemize}
    \itemindent=-13pt
    \item \textbf{Type-1 (T1):} syntactically identical code segments, with the exception of variations in whitespace, layout, and comments \cite{bellon2007comparison}.
    \item \textbf{Type-2 (T2):} syntactically identical code segments, with the sole distinctions being in identifier names and literal values, alongside the variances characteristic of Type-1 clones, such as those in whitespace, layout, and comments \cite{bellon2007comparison}.
    \item \textbf{Type-3 (T3):} syntactically similar code segments, exhibiting differences at the statement level. These segments demonstrate variations wherein statements are added, modified, and/or removed in relation to each other, in addition to the distinctions present in both Type-1 and Type-2 clones, such as disparities in whitespace, layout, comments, identifier names, and literal values \cite{roy2007survey}.
    \item \textbf{Type-4 (T4):} while syntactically dissimilar, code segments implement identical functionality \cite{roy2007survey}.
\end{itemize}

Basically, the definition of Type-1 clone and Type-2 clone, known as syntactic code clones, are pretty clear, grounded in the textual similarity of code fragments. This phenomenon is commonly observed in practices involving copying and pasting. Conversely, the classification of Type-3 and Type-4 clones, also referred to as semantic code clones, is always controversial. Distinct from syntactic clones, which are identified based on textual similarity, semantic clones may exhibit similar functionality but are often implemented using diverse syntactic structures.

In this study, we adopt the categorization framework of BigCloneBench (BCB) for code clones, which is predicated on their percentage of similarity. Clones are classified as Very-Strongly Type-3 (VST3) when they exhibit a similarity ranging from 90\% (inclusive) to 100\%. Clones falling within the 70-90\% similarity bracket are categorized as Strongly Type-3 (ST3), those within 50-70\% as Moderately Type-3 (MT3), and clones with a similarity percentage between 0-50\% are designated as Weakly Type-3 or Type-4 (WT3/4).

\subsection{Generative Pretrained Transformers (GPTs)}

With the success of Transformer-based LLMs, the field of natural language processing (NLP) has witnessed remarkable achievements. Brown et al. \cite{brown2020language} introduce the GPT-3.5 model, this model, boasting 175 billion parameters, demonstrated the potential of scaling up parameters in auto-regressive language models for improved performance across a variety of NLP tasks.

Following this, GPT-3.5 emerged as a refined iteration of its predecessor. Developed by Ouyang et al.\cite{ouyang2022training}, GPT-3.5 incorporated reinforcement learning from human feedback, enhancing its capabilities and efficiency.

GPT-4, as introduced by OpenAI \cite{achiam2023gpt} in 2023, assumed the mantle as the successor to GPT-3.5. Although its exact parameter count was not disclosed, it is presumed to be larger and more sophisticated than its predecessors. GPT-4's evaluation encompassed a diverse array of many NLP datasets, covering tasks from Question Answering (QA) and Natural Language Inference (NLI) to Reading Comprehension (RC). Uniquely, GPT-4 was also tested on sets of exams originally designed for humans, showing its advanced understanding and response capabilities.

\begin{figure*}[htbp]
\centering
    \includegraphics[width =1.0\linewidth]{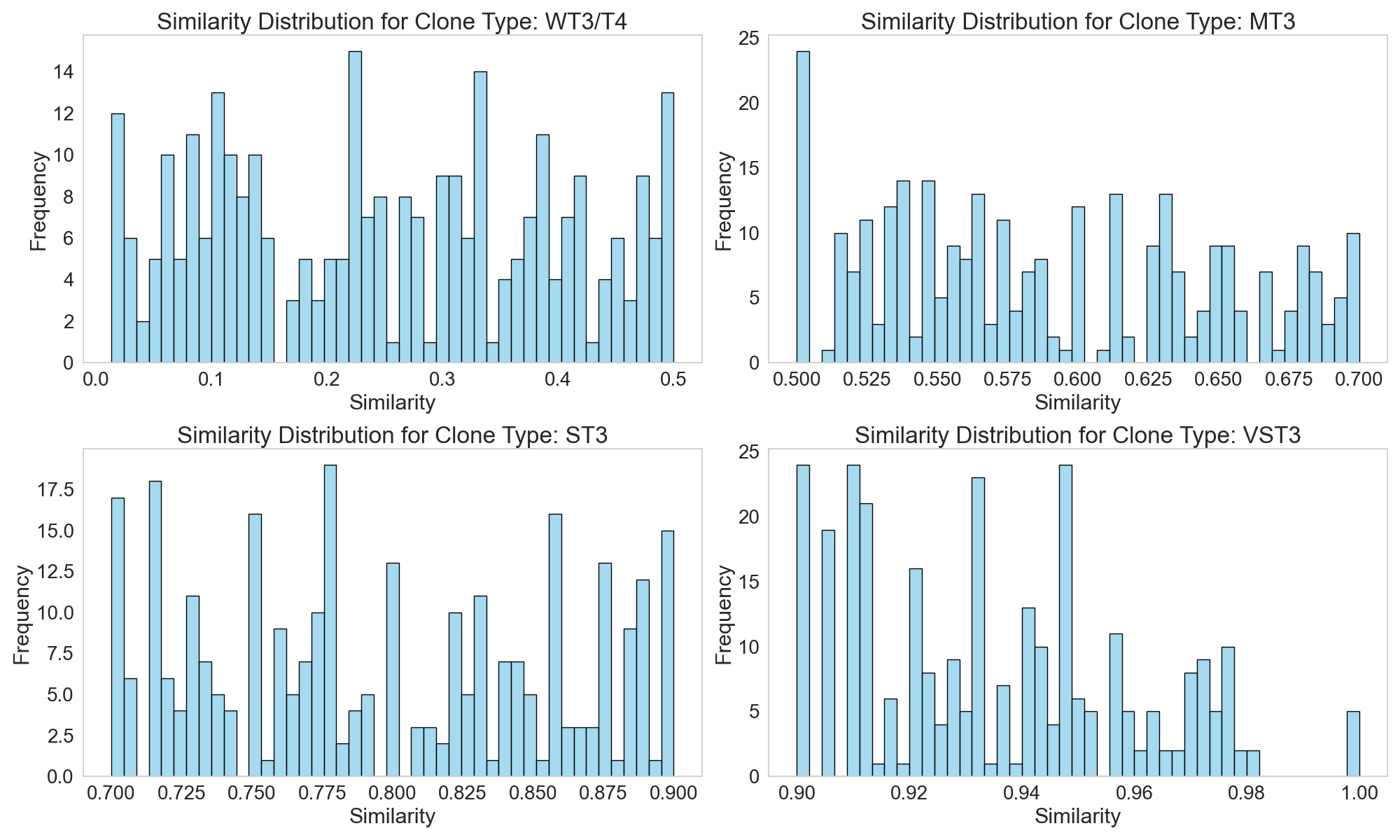}
\caption{BCB Similarity Distribution per Clone Type}
\label{fig:similarity_distributes}
\end{figure*}

\section{Approach}
\label{sec:approach}

In this section we describe our proposed approach to evaluate the performance of LLM models at identifying code clones in two parts: Code Clone Dataset, and GPT Prompt Engineering.

\subsection{Datasets}
In our study, we use two particular datasets: (i) BigCloneBench is the most used in the literature and combines human-engineered code clones, and (ii) a new dataset that combines GPT-engineered code clones.

\subsubsection{BigCloneBench \cite{10.1109/icsme.2014.77}}
A widely-used clone detection benchmark in code clone detection tasks. It is a comprehensive collection of 8 million validated clones within IJaDataset-2.0 \cite{IJaDataset2}, a repository containing 25,000 open-source Java systems. This benchmark covers both intra-project and inter-project clones across four primary clone types, spanning the entire range of clone syntactical similarity. 


\subsubsection{GPTCloneBench \cite{10.48550/arxiv.2308.13963}}
A comprehensive benchmark designed to evaluate semantic and cross-language code clones using GPT-3.5 \cite{brown2020language} and SemanticCloneBench \cite{al2020semanticclonebench}. GPTCloneBench leverages GPT-3.5's capabilities to generate semantic and cross-language clones from code fragments in SemanticCloneBench. GPTCloneBench includes several true semantic clone pairs, false semantic pairs, and cross-language clones across four programming languages (Java, C, C\#, and Python).

\subsubsection{Data Selection}

To maximize the efficiency of evaluating GPT models' performance in code clone detection while minimizing experimental costs (specifically, the usage of API tokens), we have implemented a meticulous clone selection process before their analysis with GPT models. Within the constraints of our budget, we chose 300 samples for input into the GPT models for each dataset and clone type. For the BCB dataset, which comprises multiple open-source Java repositories and contains numerous replicated code clones, our initial step involves filtering out these replicated clones to ensure the uniqueness of the code clones submitted to the GPT models. Subsequently, for each clone type, we endeavor to maintain representative code pairs with diverse similarities across the 300 code pairs, aiming for a balanced representation of clone variations.

In the case of GPTCloneBench, which serves as a benchmark for comparing the performance of GPT models between LLM-generated and human-made code clones, we employ the same selection process as with BCB. Following this, for each clone type within GPTCloneBench, we choose 300 samples to align the code size distribution for each clone type group as closely as possible with the code size distribution of the corresponding clone type in the selected samples from the BCB dataset.

The details of the selected data are presented in Table \ref{tab:data_describe}. Owing to the scarcity of clone types T1, T2, VST3, and ST3 in the GPTCloneBench dataset, our selection was limited to 300 samples each for MT3 and WT3/T4 types from this dataset. Additionally, it is important to note that both datasets are devoid of negative samples, which are non-clone pairs. Consequently, this study solely encompasses positive samples, necessitating the use of True Positive, False Negative, and Recall as the primary evaluation metrics.

\begin{table}[]
\caption{Clone Pair Summary}
\resizebox{\columnwidth}{!}{%
\begin{tabular}{lllllll}
\hline
Dataset/Clone Type & T1  & T2  & VST3 & ST3 & MT3 & WT3/T4 \\ \hline
BCB      & 300 & 300 & 300  & 300 & 300 & 300    \\ \hline
GPTCloneBench      & /   & /   & /    & /   & 300 & 300    \\ \hline
\end{tabular}%
}
\label{tab:data_describe}
\end{table}

\subsubsection{Selected Data Profile}

The distribution of code similarity for clone types WT3/T4, MT3, ST3, and VST3 in the selected data from BCB is illustrated in Figure \ref{fig:similarity_distributes}. This figure reveals that the data selected from BCB for these types contains diverse similarities, signifying that the selected data is apt for reflecting GPT models' performance across a diverse range of code pair similarities.

\begin{figure*}[htbp]
\centering
    \includegraphics[width =1.\linewidth]{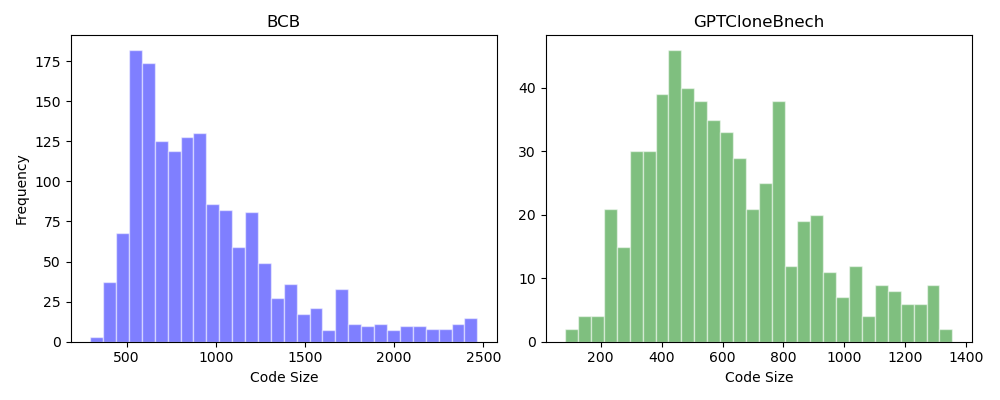}
\caption{BCB and GPTCloneBench Code Size Distributions}
\label{fig:code_size_distributes}
\end{figure*}

The Program size distributions for the data selected from BCB and GPTCloneBench are depicted in Figure~\ref{fig:code_size_distributes}. The bulk of the codes fall within the size ranges 400-600 and 250-750 Bytes for BCB and GPTCloneBench respectively.

\subsection{GPT Prompt Engineering}
A clear and accurate prompt is essential in the execution of natural language processing tasks. A prompt constitutes a set of input instructions or guidelines aimed at eliciting a specific output or accomplishing a particular task. It may take the form of a question, a statement, or an instruction. Prompts play a pivotal role in facilitating interaction with LLMs, thereby aiding in achieving the intended outcome from the model's response. 

There exist various techniques to define prompts, (e.g., zero-shot~\cite{palatucci2009zero, wu2019zero}, one-shot~\cite{fei2006one}, few-shot~\cite{fe2003bayesian}, chain of thought~\cite{wei2022chain}, self-improving~\cite{chen2017learning}, or analogy-based~\cite{hofstadter2001analogy}). Zero-shot employs direct instructions without the need for pre-training data or prior knowledge input. One-shot enables a model to perform tasks based on a single example. Few-shot trains a model using only a handful of examples for new tasks. Chain of thought prompts encourage models to detail their reasoning step by step. Self-improving prompts guide models to critique and enhance their outputs over time. Analogy-based prompts help models solve problems by drawing comparisons to familiar scenarios. Various prompt techniques are suitable for different situations. The selection of a prompt can substantially influence the performance of the model.

\subsubsection{Impractical One-Shot Prompt}
We first analyzed the feasibility of a one-shot prompt using this prompt: \emph{``Determine whether two provided code snippets are code clones. Output yes or no with no explanation.''}.

We conducted a manual evaluation of the one-shot prompt technique using several code pair samples. Our findings indicate that the GPT models do not yield consistent results from this one-shot prompt approach, showing whether the model provides explanations for its judgment or exhibits misunderstandings regarding code clones. 

\subsubsection{Selecting Few-Shot Prompt}
\begin{figure}[htbp]
\centering
    \includegraphics[width=1.0\linewidth]{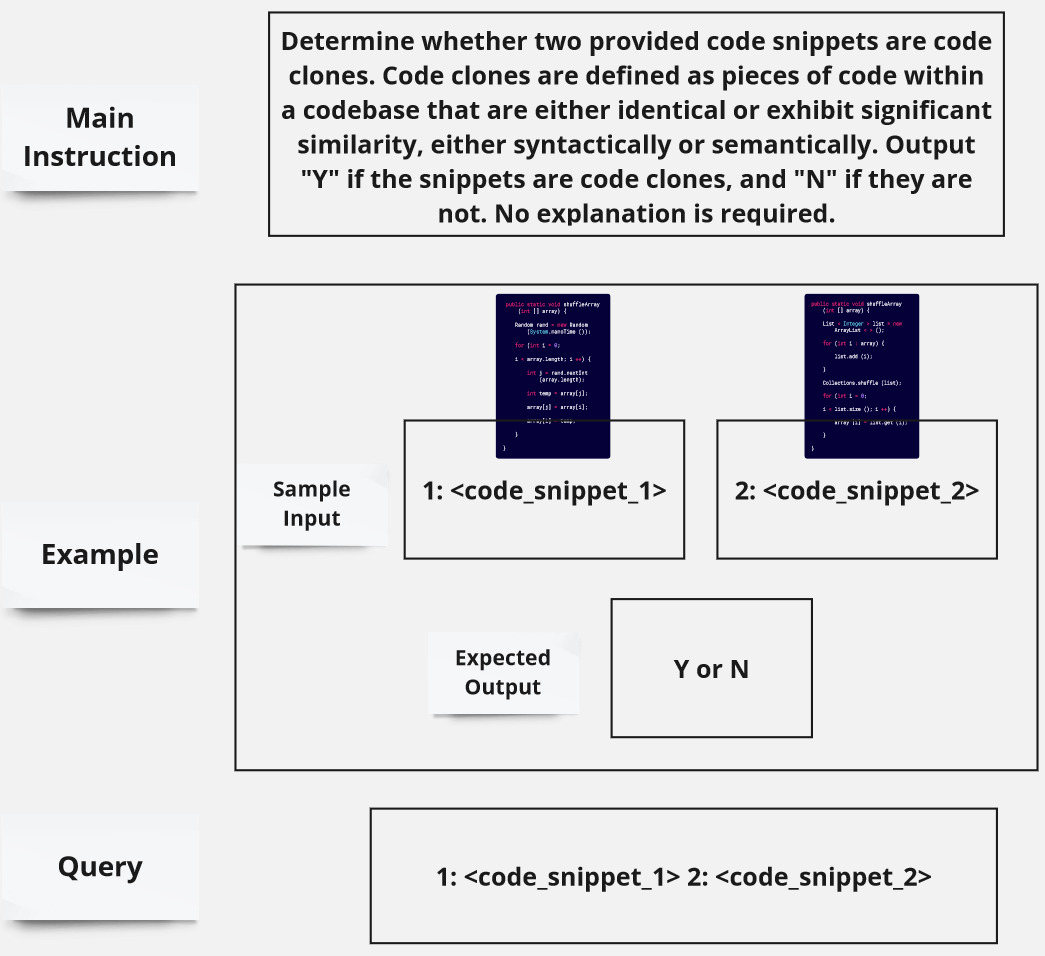}
\caption{Few-Shot GPT Prompt}
\label{fig:prompt}
\end{figure}

We shifted our approach to a few-shot prompt technique to accurately assess and contrast the performance of the GPT models. Specifically, we constructed our prompt with simple instructions, an input sample, and definitions of code clones to direct GPT models in determining whether a given code pair is a clone or not. Figure~\ref{fig:prompt} shows an example of the creation of the prompt with a code pair. The prompt is segmented into three distinct parts. The first section comprises the main instruction, which outlines the basic task description and provides a definition of code clones to the GPT model. The second section offers an example, supplying GPT with a sample input and its corresponding output. The final section poses a query to the GPT model, featuring two code snippets labeled with numbers and colons, requesting the GPT to render a determination on whether the specified code pair is a clone. By employing multiple code pair samples, we find that this few-shot prompt yields stable and expected results from the GPT model.

\begin{figure*}[htbp]
\centering
    \includegraphics[width=.9\linewidth]{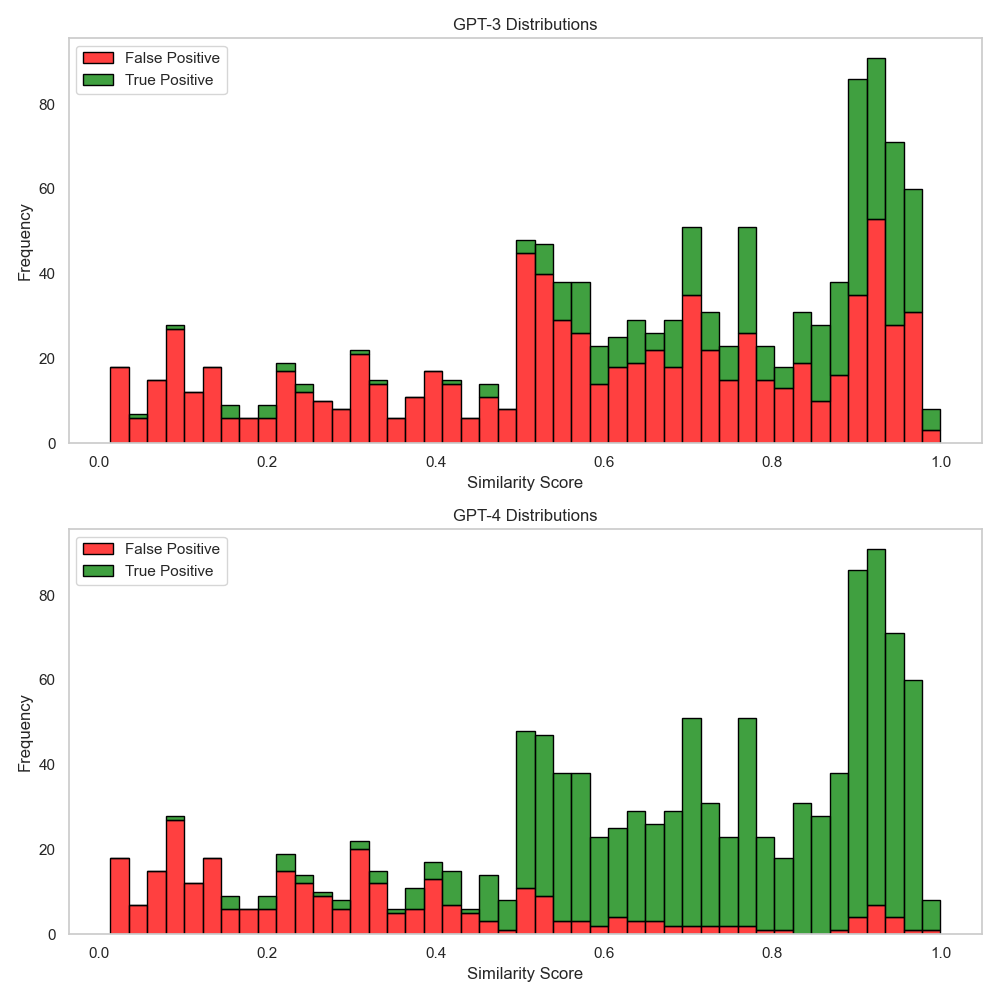}
\caption{Performance of LLM models per Code Pair Similarity}
\label{fig:gpt_answer_similarity_distribution}
\end{figure*}

\section{Results}
\label{sec:results}

In this section, we answer the two research questions defined in Section~\ref{sec:introduction} based on our experimental analysis.



\subsection{LLM Performance at Detecting Clone Clones}

In our initial comparison of code clone detection performance between GPT-3.5 and GPT-4, we focused on the distributions of true positive and false negative answers across four different Clone Types, utilizing data selected from BCB. The findings are presented in Table \ref{table:gpt_answer_clone_types}. For clone Type T1, both GPT-3.5 and GPT-4 achieved a 100\% rate of true positives with zero false negatives, demonstrating their proficiency in recognizing Type-1 clones. In the case of clone Type T2, GPT-3.5 exhibited a lower recall of 0.56 in contrast to GPT-4, which achieved 0.86, indicating a notably higher success rate for GPT-4 with this clone type.

For the more semantically complex clone types, VST3, ST3, and MT3, the performance of the GPT-3.5 model declined progressively, with recall generally falling below 0.50. Conversely, the GPT-4 model displayed significantly superior performance, with all recalls exceeding 0.85. For clone Type-4, characterized by pair similarities of less than 50\%, both models showed relatively low performance levels; however, GPT-4 still outperformed GPT-3.5 with a recall of 0.23.


Our analysis further extends to comparing GPT-3.5 and GPT-4 in terms of their performance relative to the distribution of code pair similarity, as detailed in Figure \ref{fig:gpt_answer_similarity_distribution}. We notice a pronounced increase in the frequency of true positives within the higher similarity score range (notably from 0.5 to 1.0), corresponding to clone Type-3. This observation suggests a more substantial accumulation of true positives within this range for GPT-4 compared to GPT-3.5, hinting at GPT-4's enhanced capability in accurately identifying code clones with relatively higher similarity.

In the segment pertaining to clone Type-4, where the similarity score trends towards 0.5, the data indicates a predominance of false positives over true positives for both models. Nonetheless, there is an observable decline in the frequency of positive identifications as the similarity score diminishes towards 0.5. Despite this, GPT-4's performance remains superior to that of GPT-3.5 in this specific region, underscoring its improved effectiveness in distinguishing code clones, even within lower similarity scores. 

\textit{\textbf{Answer to RQ1: }} The study finds that GPT-4 consistently outperforms GPT-3.5 across all clone types, particularly for code pairs exhibiting low similarity, highlighting GPT-4's advanced ability to understand semantic nuances within the code.

\begin{table}[]
\begin{center}
\caption{Performance of LLM Models on Different Clone Types}
\label{table:gpt_answer_clone_types}
\resizebox{\columnwidth}{!}{%
\begin{tabular}{|l|l|p{1.5cm}|p{1.5cm}|p{1.5cm}|}
\hline
Clone Type            & Model   & TP  & FP  & Recall \\ \hline
\multirow{2}{*}{T1}   & GPT-3.5 & 300 & 0   & 1.00   \\ \cline{2-5} 
                      & GPT-4   & 300 & 0   & 1.00   \\ \hline
\multirow{2}{*}{T2}   & GPT-3.5 & 169 & 131 & 0.56   \\ \cline{2-5} 
                      & GPT-4   & 259 & 41  & 0.86   \\ \hline
\multirow{2}{*}{VST3} & GPT-3.5 & 156 & 144 & 0.52   \\ \cline{2-5} 
                      & GPT-4   & 283 & 17  & 0.94   \\ \hline
\multirow{2}{*}{ST3}  & GPT-3.5 & 133 & 167 & 0.44   \\ \cline{2-5} 
                      & GPT-4   & 290 & 10  & 0.97   \\ \hline
\multirow{2}{*}{MT3}  & GPT-3.5 & 70  & 230 & 0.23   \\ \cline{2-5} 
                      & GPT-4   & 262 & 38  & 0.87   \\ \hline
\multirow{2}{*}{WT3/T4}   & GPT-3.5 & 20  & 280 & 0.07   \\ \cline{2-5} 
                      & GPT-4   & 68  & 232 & 0.23   \\ \hline
\end{tabular}%
}
\end{center}
\end{table}

\subsection{Difference Between Human-Made and LLM-Generated Code Clones}

\begin{table}[]
\begin{center}
\caption{Performance of LLM models on BCB (Human-Generated) and GPTCloneBench (LLM-Generated)}
\label{table:table_q2}
\resizebox{\columnwidth}{!}{%
\begin{tabular}{|l|l|l|l|l|l|}

\hline
Clone Type           & Model                    & Data Source   & TP  & FP  & Recall \\ \hline
\multirow{4}{*}{MT3} & \multirow{2}{*}{GPT-3.5} & BCB           & 70  & 230 & 0.23   \\ \cline{3-6} 
                     &                          & GPTCloneBench & 207 & 93  & 0.69   \\ \cline{2-6} 
                     & \multirow{2}{*}{GPT-4}   & BCB           & 262 & 38  & 0.87   \\ \cline{3-6} 
                     &                          & GPTCloneBench & 252 & 48  & 0.84   \\ \hline
\multirow{4}{*}{WT3/T4}  & \multirow{2}{*}{GPT-3.5} & BCB           & 20  & 280 & 0.07   \\ \cline{3-6} 
                     &                          & GPTCloneBench & 174 & 126 & 0.58   \\ \cline{2-6} 
                     & \multirow{2}{*}{GPT-4}   & BCB           & 68  & 232 & 0.28   \\ \cline{3-6} 
                     &                          & GPTCloneBench & 232 & 68  & 0.77   \\ \hline
\end{tabular}%
}

\end{center}
\end{table}

Table \ref{table:table_q2} presents the comparative analysis results from different data sources across various clone types (MT3 and WT3/T4) and GPT models (GPT-3.5 and GPT-4). The data sources in question include BCB, which consists of human-generated clones, and GPTCloneBench, comprising LLM-generated clones. The results demonstrate that the LLM-generated GPTCloneBench yields superior performance in terms of higher recall rates compared to the human-generated BCB data across both clone types. This suggests that GPT models exhibit enhanced proficiency in recognizing code clones that they themselves have generated, as opposed to real-world code clones. Nonetheless, for GPT-4, the difference in performance between BCB and GPTCloneBench is less marked in recognizing MT3 code clones, indicating its robustness in clone detection across sources.

Despite the disparities in evaluation data, we have documented the performance metrics of learning-based code clone detection algorithms for comparison. For reference, existing code clone detection techniques achieve recall rates of 0.94 and 0.45 for clone Type-4 on the BCB~\cite{zhang2023efficient} and  GPTCloneBench~\cite{saini2018oreo} datasets, respectively. These references provide further evidence of a bias within GPT models towards more accurately identifying code clones that they themselves have generated, as opposed to recognizing clones originating from real-world scenarios.

\textit{\textbf{Answer to RQ2: }} The findings indicate that both models perform better on LLM-generated clones compared to human-generated clones, with GPT-4 showing less performance difference between the two sources, suggesting its robustness in clone detection across sources.

\section{Conclusion and Future work}
\label{sec:conclusion}

While several advanced ML techniques have been proposed to detect code clones, no work has, so far, employed LLMs for that purpose. In this study, we explored the capabilities of GPT-3.5 and GPT-4 in detecting code clones, utilizing datasets of both human-made and LLM-generated clones. 

Our experiments show a superior performance GPT-4 over GPT-3.5 regardless of clone types and data sources. Notably, both models demonstrated lower effectiveness in identifying semantic (Type-4) clones, which presents a higher challenge in enabling LLMs to grasp semantic information in code. Additionally, LLM-generated clones were detected with greater accuracy compared to human-made clones, suggesting an intrinsic model bias toward recognizing familiar patterns.

In our future work, we will implement an evaluation on a larger scale, with a higher amounts of code clones and a variety of programming languages within open-source LLMs to get rid of budget constraints. Furthermore, we plan to compare the performance of LLMs with other state-of-art code clone detection tolls, as well as traditional code clone detection tools, to provide a comprehensive analysis of LLM's performance in code clone detection tasks.

\section*{Acknowledgment}
This research was supported by Science Foundation Ireland under grant numbers 18/CRT/6223 (SFI Centre for Research Training in Artificial Intelligence), and 13/RC/2094/P\_2 (Lero SFI Centre for Software). For the purpose of Open Access, the author has applied a CC BY public copyright license to any Author Accepted Manuscript version arising from this submission.

\bibliographystyle{ACM-Reference-Format}
\bibliography{ref}

\end{document}